\documentclass[prc,superscriptaddress,showpacs,twocolumn,notitlepage]{revtex4-1} 
\usepackage{amssymb}
\usepackage{amsmath}
\usepackage[usenames]{color}
\usepackage{graphicx}
\usepackage{soul}
\usepackage{pifont}
\usepackage{bbm}
\usepackage{multibib}
\usepackage[colorlinks]{hyperref}
 
\allowdisplaybreaks
\newcommand{\ve}[1]{\ensuremath{\boldsymbol{#1}}}

\newcommand{\ket}[1]{\vert #1 \rangle}

\begin{document}
\title{Majorana quasiparticles in semiconducting carbon nanotubes}

\author{Magdalena Marganska}
\affiliation{Institute for Theoretical Physics, University of Regensburg, 93053 
Regensburg, Germany} 
\author{Lars Milz}
\affiliation{Institute for Theoretical Physics, University of Regensburg, 93053 
Regensburg, Germany} 
\author{Wataru Izumida}
\affiliation{Institute for Theoretical Physics, University of Regensburg, 93053 
Regensburg, Germany} 
\affiliation{Department of Physics, Tohoku University, Sendai 980 8578, 
Japan} 
\author{Christoph Strunk}
\affiliation{Institute for Experimental and Applied Physics, University of Regensburg, 93053 
Regensburg, Germany} 
\author{Milena Grifoni}
\thanks{Corresponding author. Email: milena.grifoni@ur.de}
\affiliation{Institute for Theoretical Physics, University of Regensburg, 93053 
Regensburg, Germany} 


\begin{abstract}
Engineering effective p-wave superconductors hosting Majorana quasiparticles (MQPs) is nowadays of
particular interest, also in view of the possible utilization of MQPs in fault-tolerant topological quantum computation.
In quasi one-dimensional systems, the parameter space for topological superconductivity is significantly
reduced by the coupling between transverse modes.
Together with the requirement of achieving the topological phase under experimentally feasible conditions, this strongly restricts in practice the
choice of systems which can host MQPs.  
Here we demonstrate that semiconducting carbon nanotubes (CNTs) in proximity with ultrathin s-wave superconductors,
e.g. exfoliated NbSe$_2$, satisfy these needs. By precise numerical tight-binding calculations in the real space we
show the emergence of localized zero-energy states at the CNT ends above a critical value of the applied magnetic field.
Knowing the microscopic wave functions, we unequivocally demonstrate the Majorana nature of the localized states.
An accurate analytical model Hamiltonian is used to calculate the topological phase diagram.
\end{abstract}
\maketitle
\section{Introduction}
\begin{figure}[tbp]
\includegraphics[width=\columnwidth]{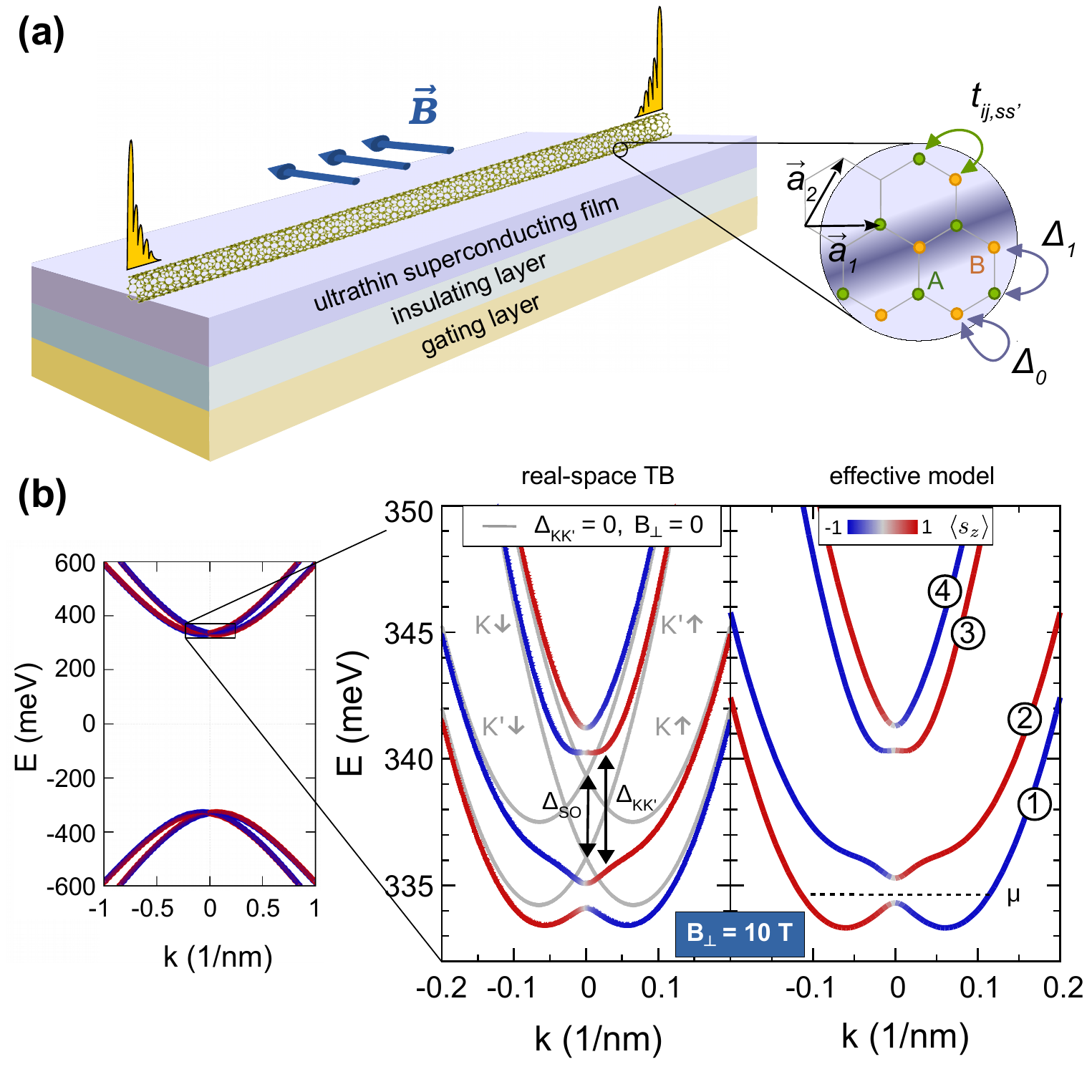} 
\caption{\label{fig:setup-bulk}
Set-up and bulk properties of a proximitized nanotube.
(a) Schematic of the system, the CNT with its proximal superconductor and a gating layer. A 
magnetic field is applied in parallel to the substrate and perpendicular to the nanotube.
We find Majorana quasiparticles at the ends of the CNT/superconductor hybrid. The ingredients
of our model are shown in the inset. The nearest neighbor hopping $t_{ij,ss'}$ 
is spin-dependent because of spin-orbit coupling. The superconducting substrate (i) breaks the
rotational symmetry of the nanotube, as shown by the darker strip with finite electrostatic on-site 
potential, and (ii) induces superconducting pairing in the nanotube, with on-site
($\Delta_0$) and nearest-neighbor ($\Delta_1$) pairing correlations.
(b) The energy bands of a (12,4) nanotube in the vicinity of the Dirac points 
are shown in the leftmost plot, with red/blue corresponding to spin up/down (quantized along
the nanotube axis) bands. Our region of interest here is the neighborhood of the $\Gamma$ point
in the conduction band. The enlarged plots show the spectrum in this region, obtained both in the
real-space tight-binding calculation and in an analytical effective model. The spin-orbit splitting
between the Kramers doublets at $k=0,B_\perp=0$ is $\Delta_{\mathrm{SO}}$ (here equal 2~meV), and the width of the 
anticrossing opening between different valley states is $\Delta_{KK'}$ (here 2.5~meV). Grey lines shown in the 
plot correspond to subbands without the valley mixing. There we can assign spin and 
valley quantum number to each band. With the valley mixing,
$B_\perp$ is able to open a gap at $k=0$. 
}
\end{figure}

Majorana fermions, particles being their own antiparticle  predicted  already eighty years ago \cite{majorana:inc1937}, have remained elusive to experimental observation so far. Hence, recent proposals to observe quasiparticles with the Majorana property - so called Majorana quasiparticles (MQPs) - in one-dimensional (1D) hybrid systems containing superconducting elements~\cite{kitaev:condmat2000} have raised big attention.
The most popular implementations are based on semiconducting nanowires with large spin-orbit interaction and large g-factor,  proximity coupled to a conventional superconductor  \cite{lutchyn:prl2010,oreg:prl2010}. When a magnetic field is applied to the nanowire in the direction perpendicular to the effective spin-orbit field,  a topologically non trivial phase is expected when the induced  Zeeman  splitting  is large enough to overcome the superconducting gap. Signatures of MQP behavior include e.g. a quantized zero-bias peak emerging in transport spectra while sweeping the magnetic field.     
Setups  with epitaxially grown superconductor-semiconducting nanowires are by now the most  advanced experimentally, and the emergence  of a zero bias transport peak at finite magnetic field 
 has been reported by various groups \cite{mourik:science2012,deng:nanolett2012,churchill:prb2013,deng:science2016}. 
Zero-bias peaks can however also emerge due to the coalescence of Andreev bound states \cite{deng:science2016,liu:prb2017} - naturally occurring in confined normal conductor-superconductor systems - or due to the development of Kondo correlations \cite{lee:prl2012}. 
An unambiguous theoretical confirmation of the experimental observation of MQPs would require an accurate microscopic modeling of the nanowires.
However, diameters of many tens of nanometers and lengths of several micrometers hinder truly microscopic calculations of the electronic spectrum of finite systems.
The real space models of semiconductor nanowires are usually constructed in a top-down approach, starting with an effective model and quantizing it on a chosen crystal lattice~\cite{stanescu:jpcm2013}. 
Without accurate modelling of experimental set-ups one can make only qualitative, rather than quantitative predictions of the boundaries of the topological phase. Recently MQP signatures have also been observed in Kitaev chains of magnetic adatoms chains on superconducting substrates \cite{nadj-perge:science2014,ruby:prl2015}. The microscopic modeling of ferromagnetic chains is however still in development~\cite{peng:prl2015,pawlak:npjqi2016}. 
%
In this work we consider carbon nanotubes (CNTs) as host for MQPs. Due to their small diameter, they can be considered as truly 1D conductors with one relevant transverse mode for each valley and spin. 
The low energy spectrum of the CNTs is well described in terms of tight-binding models for carbon atoms on a rolled graphene lattice~\cite{saito:1998}. Experimental advances in the preparation of ultraclean CNTs have allowed to measure their transport spectra in various transport regimes \cite{laird:rmp2015}, and hence to gain confidence in the accuracy of the theoretical modeling. Two proposals to observe MQPs in carbon nanotubes have been based on spiral magnetic fields~\cite{egger:prb2012}, induced e.g. by magnetic domains~\cite{kontos:private},  or on large electrical fields \cite{klinovaja:prl2012}. Despite their appeal due to the possibility of inducing large extrinsic spin-orbit coupling, these set-ups are quite sophisticated and either hard to realize experimentally or to model microscopically.\\
The set-up which we describe here is, similar to \cite{sau:prb2013}, based solely on the intrinsic curvature-induced spin orbit coupling of CNTs.
It consists of a CNT placed on an ultrathin superconducting film, with a gating layer beneath and the magnetic field applied parallel to the film
and perpendicular to the nanotube, shown in Fig.~\ref{fig:setup-bulk}a). 
The small size of CNTs allows us to use a bottom-up approach starting with a tight-binding model of the CNT lattice, with external influences such as the substrate potential, superconducting pairing and magnetic field added in the real space. We then construct effective Hamiltonians in the reciprocal space which well reproduce the numerically calculated low energy spectrum. This in turn allows us to gain the knowledge of the system's symmetries and topological invariants.
In this work we consider semiconducting rather than the metallic CNTs proposed in \cite{sau:prb2013}, since the Fermi velocity in the former is lower by a factor of $\sim 10^{-3}$ than in the latter. 
We can thus achieve energy quantization sufficient to resolve the subgap features in much shorter CNTs.
In consequence, semiconducting CNTs can host Majorana end states at a thousand times smaller length than the metallic ones,
which improves the experimental chances of synthesizing an ultraclean device. As we shall show, MQPs can arise at the end of proximitized CNTs with length of only a few micrometers, easily handled in the device synthesis.\\

\section{Set-up and bulk properties}
Carbon nanotubes can be regarded as graphene sheets rolled into seamless cylinders. The rolling 
direction is described by the so-called chiral indices of the CNT, $(n,m)$~\cite{saito:1998}. The bulk spectrum of the 
CNT consists of 1D subbands created by cutting graphene's dispersion by lines of constant angular
momentum, determined by the periodic boundary conditions around the circumference.
The electronic properties of the nanotube depend strongly on the rolling direction, which
decides whether the subbands cross the Dirac points or not. If they do, i.e.
when $(n-m)|\mathrm{mod}\,3 = 0$ the nanotubes are metallic and the lowest 1D conduction bands descend deeply
towards the apex of graphene's Dirac cones, reaching Fermi velocities of the order of $10^6$~m/s. 
If $(n-m)|\mathrm{mod}\,3 \neq 0$, the CNT is semiconducting and the lowest bands lie higher up on 
the Dirac cones and are much flatter, with Fermi velocities dependent on the chemical potential, but
typically not higher than $\sim10^3$~m/s. In the following we
shall use for illustration a finite (12,4) CNT, although we find the same topological phases in
semiconducting nanotubes of other chiralities, in different parameter regimes.\\
The microscopic model of the nanotube which we use, with one $p_z$ orbital per atomic site,
is shown schematically in Fig.~\ref{fig:setup-bulk}(a).
The tiny spin-orbit coupling of graphene becomes significantly enhanced in carbon nanotubes
due to the curvature of their atomic lattice~\cite{ando:jpsj2000,kuemmeth:nature2008,izumida:jpsj2009}. 
It defines a quantization axis for the spin, along the CNT axis, and induces a band splitting $\Delta_{SO}$, which 
is reported to reach values larger than 3~meV~\cite{steele:natcomms2013}.
The resulting low energy band structure for a (12,4) semiconducting nanotube is shown in the small panel of 
Fig.~\ref{fig:setup-bulk}b) and, zoomed up around the $\Gamma$ point, with the grey lines in the larger panel.
The band crossing at $k=0$ is protected by
symmetry since the crossing bands belong to different valleys $K$ and $K'$, i.e. in this CNT to different
angular momenta~\cite{marganska:prb2015,izumida:prb2016}, and the magnetic field cannot hybridize them. 
The presence of a superconducting substrate plays here a double role. On the one hand it serves as a source of superconducting correlations
in the nanotube, acquired by the proximity effect.
On the other hand it breaks the rotational symmetry of the nanotube
and is the cause of valley mixing $\Delta_{KK'}$. In combination with the perpendicular magnetic field $B_\perp$, this allows 
the bands at the $\Gamma$ point to hybridize.
The increased electrostatic potential in the vicinity of the substrate atoms is shown
as a darker stripe across the inset in Fig.~\ref{fig:setup-bulk}(a). The real space CNT
Hamiltonian in the presence of perpendicular magnetic field $B_\perp$ is then given by
\begin{equation}
\label{eq:single-particle}
\begin{split}
H_0 & = \sum_{\langle i,j\rangle, ss'} t_{ij,ss'} c_{is}^\dag c_{js'}^{\vphantom{\dag}} +  \sum_{i,s} V(\varphi_i) c_{is}^\dag c_{is}^{\vphantom{\dag}} \\
 &+ \mu_B B_\perp \sum_{i,s} c_{is}^\dag c_{i,-s}^{\vphantom{\dag}},
 \end{split}
\end{equation}
where $i$ indexes the atomic positions, $s$ is the spin, $t_{ij,ss'}$ is the spin-dependent
nearest neighbor hopping~\cite{ando:jpsj2000},
$\langle i,j\rangle$ denotes a sum over the nearest neighbor atoms, and $V(\varphi_i)$ is the 
potential induced by the substrate at the $i$-th nanotube atom. It depends on the atom's height above the substrate,
i.e. on its angular coordinate $\varphi_i$.
The resulting band structure is shown in the left large panel of Fig.~\ref{fig:setup-bulk}(b), featuring both the
helical, spin-momentum locked modes and two energy ranges with odd number of Fermi surfaces. We have
also constructed a four-band effective model in the reciprocal space (detailed discussion follows in the Methods section), 
with the band structure shown also in Fig.~\ref{fig:setup-bulk}(b). A very good agreement with the spectrum obtained
from the full tight-binding calculation is achieved, which is crucial in the studies of topological matter.\\
When the substrate turns superconducting, it induces Cooper pairing in the nearby normal system.
We propose to use the two-dimensional (2D) gate-tunable superconductor NbSe$_2$, where superconductivity can survive up to 30~T in magnetic fields applied in-plane~\cite{xi:natphys2016}.
Hence in our set-up the magnetic field is applied in the direction perpendicular to the nanotube
axis but, crucially, parallel to the substrate.
We treat the superconducting correlations in the spirit of Ref.~\cite{uchoa:prl2007}, admitting 
both the on-site and nearest-neighbor pairing $\Delta_0$ and $\Delta_1$.
With the superconducting pairing the system is described by
\begin{eqnarray}
H &=& H_0 - \mu\sum_{i,s}c_{is}^\dag c_{is}^{\vphantom{\dag}} + \sum_{i,s} (\Delta_0 c_{is}^\dag c_{i,-s}^\dag
+ \mathrm{h.c.}) \nonumber \\
& & + \sum_{\langle i,j\rangle,s} (\Delta_1 c_{is}^\dag c_{j,-s}^\dag + \mathrm{h.c.}),
\label{eq:superconducting}
\end{eqnarray}

\begin{figure}[tbp]
\includegraphics[width=\columnwidth]{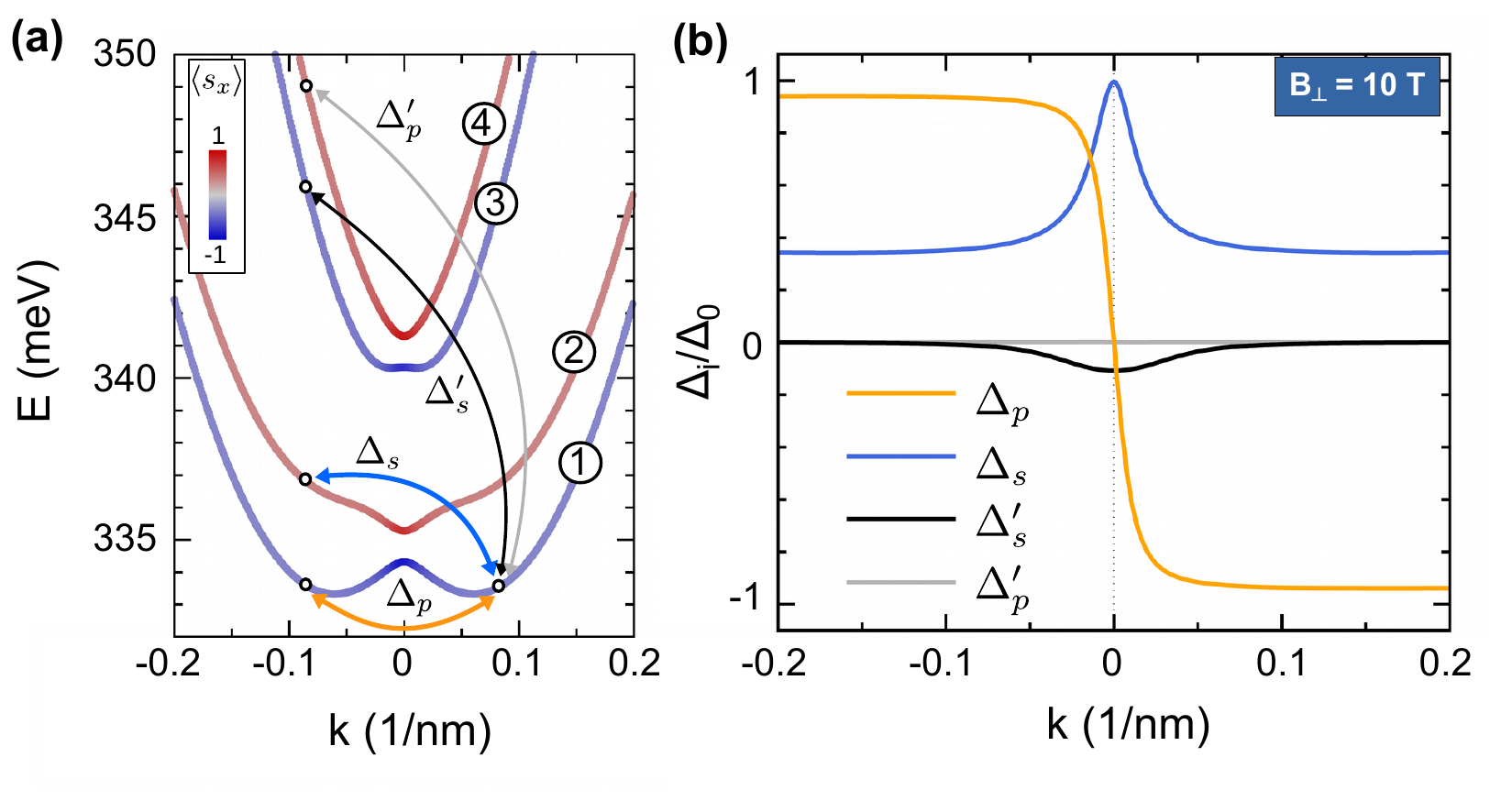} 
\caption{\label{fig:pairings}
Superconducting pairing between the four bands.
(a) The energy bands colored by their expectation value of $s_x$, spin in the direction of the field.
The superconducting pairing couples all four bands. The pairings between a positive momentum state in band \ding{192} 
and its negative momentum partners in all bands are indicated. Their strength depends on the value of $B_\perp$.
(b) The s-like pairings are even in $k$, the p-like pairings are odd. 
}
\end{figure}
\begin{figure}[tp]
\includegraphics[width=0.8\columnwidth]{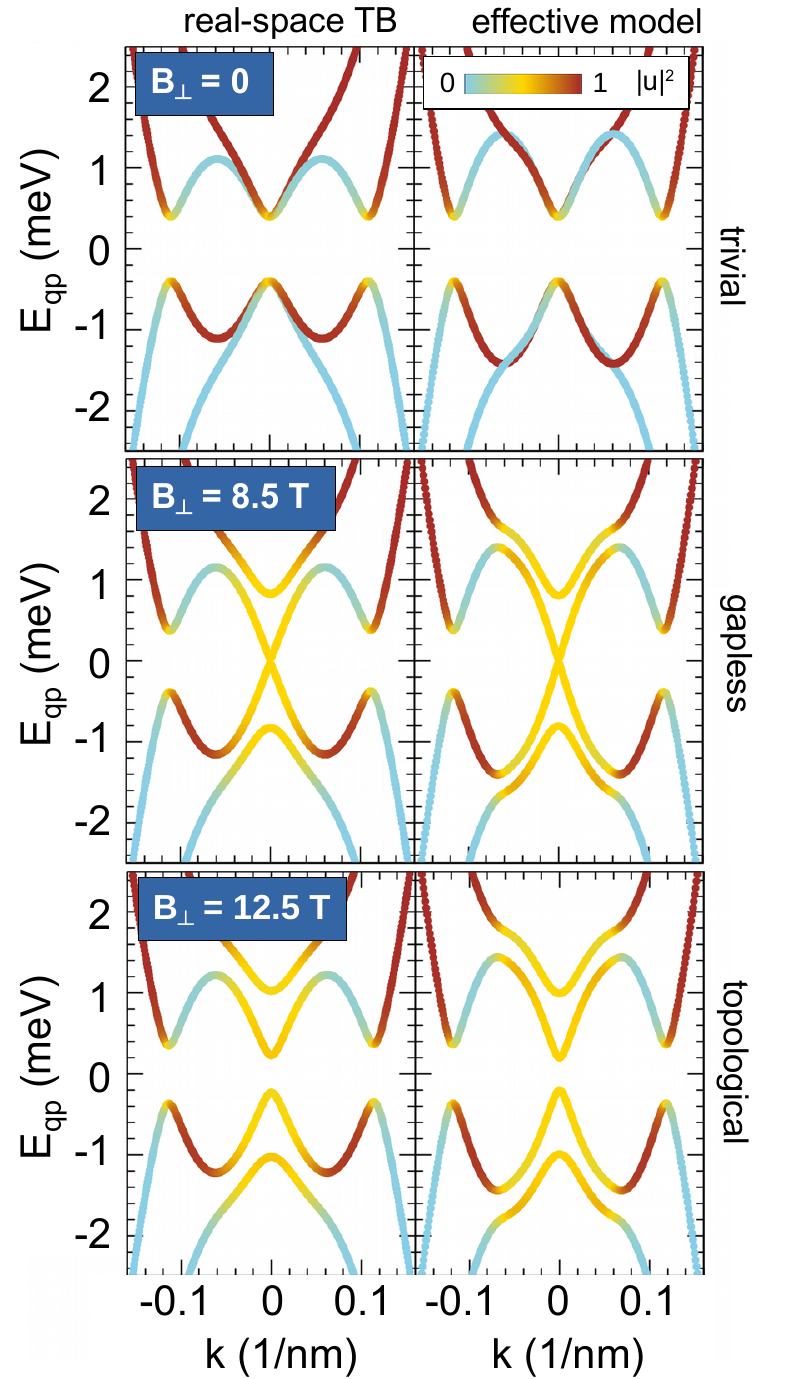}
\caption{\label{fig:BdGspectrum}
The Bogoliubov-de Gennes spectra of the superconducting nanotube in three different 
topological phases which can be accessed by tuning $B_\perp$. The color scale shows the weight of the
particle part of the corresponding CNT's eigenstate; gold color
indicates equal particle and hole contributions. The superconducting pairing is $\Delta_0=0.4$~meV, $\Delta_1=0$.
}
\end{figure}


where energies are measured from the chemical potential $\mu$, controlled e.g. by a gating layer beneath the substrate. 
The $\Delta_1$ contribution is not necessary for the MQPs to arise and we shall discuss its effects further only in
last section of the Supplemental Material, here assuming $\Delta_0\in\mathbb{R}$ and $\Delta_1 = 0$. In the effective model, whose construction
is described in the Methods section, the superconducting pairing from \eqref{eq:superconducting} couples all four single-particle bands as shown in Fig.~\ref{fig:pairings}(a), with some pairing terms having an s-wave and some a p-wave symmetry, visible in Fig.~\ref{fig:pairings}(b). \\
In order to find its spectrum, we express the Hamiltonian (\ref{eq:superconducting}) in a particle-hole symmetric form by introducing a Nambu spinor,
$\Psi = \oplus_{i=1}^N \Psi_i,\quad \Psi_i^\dag = (c_{i\uparrow}^\dag, c_{i\downarrow}^\dag, c_{i\uparrow},c_{i\downarrow})$,
where $\oplus$ is the direct sum over the $N$ atomic positions. 
\footnote{A direct sum $\ve{A}\oplus\ve{B}$ of a $p$-component vector $\ve{A}$ and an $q$-component vector $\ve{B}$ is a $p+q$-dimensional vector whose first $p$ components are those of $\ve{A}$ and the last $q$ are those of $\ve{B}$.
Our $\Psi$ and $\ve{\chi}^n$ are both $4N$-dimensional vectors. The components of $\Psi$ are operators, while those of $\ve{\chi}^n$
are complex numbers.} 
This procedure effectively doubles the number of degrees of freedom of the system. 
The full Hamiltonian becomes $H = \frac{1}{2} \Psi^\dag H_{\mathrm{BdG}} \Psi$,
where the field operators are contained in $\Psi,\Psi^\dag$ and
$H_{\mathrm{BdG}}$ is an ordinary matrix, the Bogoliubov-de Gennes Hamiltonian of our system. Its
eigenvectors, defining the quasiparticle eigenstates with a set of quantum numbers $n$, have the structure
\begin{equation}
 \label{eq:solutions}
 \ve{\chi}^n = \oplus_{i=1}^N \ve{\chi}^n_i,\quad (\ve{\chi}^n_i)^T = (u^n_{i\uparrow},u^n_{i\downarrow},v^n_{i\uparrow},v^n_{i\downarrow}),
\end{equation}
where $n$ is a generic collective index which may contain e.g. the valley and, in a system with translational invariance, $k$ quantum numbers.
The particle components with spin $s$ on atom $i$ are denoted by $u^n_{is}$ and the corresponding hole components by $v^n_{is}$. The quantum eigenstates of the system have the form $\ket{\psi^n} = \oplus_{i=1}^N \Psi_{i}^\dag\cdot\chi^n_{i}\,\ket{0}_{\mathrm{BCS}}$, where $\ket{0}_{\mathrm{BCS}}$ is the BCS ground state in the CNT. 
The low energy bands obtained for our proximitized infinite (12,4) nanotube  
are shown in Fig.~\ref{fig:BdGspectrum}, for the three topologically distinct phases encountered by increasing the magnetic field.
The color scale shows the overall weight of particle component in the given energy eigenstate, $|u|^2=\sum_{is} |u_{is}|^2$.
The solutions which have a predominantly particle character trace the original single-particle bands, while the predominantly hole-type
solutions are mirror-reflected around the chemical potential.\\

\section{Symmetries and topological invariants}
\begin{figure}[htp]
\includegraphics[width=\columnwidth]{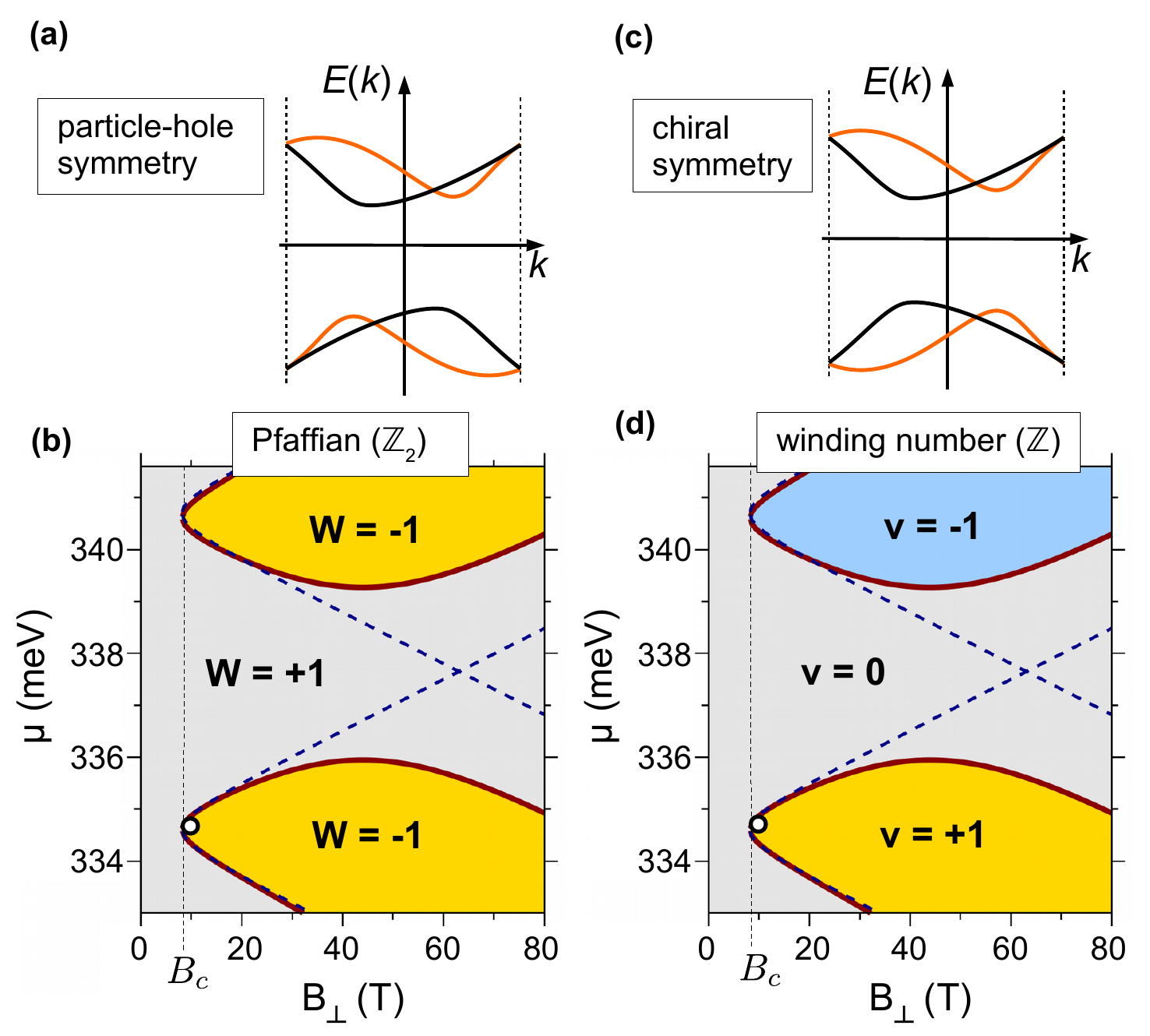}
\caption{\label{fig:invariants}
Symmetries and topological invariants.
(a) Sketch of a spectrum with particle-hole symmetry. Bands of the same color are related by the symmetry.
(b) The phase diagram calculated using the effective model and the Pfaffian formulation 
of the topological invariant, typical for particle-hole symmetric systems. The topologically 
non-trivial regions are shown in yellow, the red line at the border between the phases is 
the contour of $E=0$ at the $\Gamma$ point. The dot in the lower $W=-1$ area marks the $\mu$
and $B_\perp$ used in Fig.~\ref{fig:majoranas}. The dashed lines trace the borders of non-trivial
phase calculated from a model which contains only one single-particle band pair,
either \ding{192} and \ding{193} (higher region) or \ding{194} and \ding{195} (lower region) from Fig.~\ref{fig:setup-bulk}b.
(c) Sketch of a spectrum with chiral symmetry.
The Bogoliubov-de Gennes spectrum in Fig.~\ref{fig:BdGspectrum} has both particle-hole and chiral
symmetry. 
(d) The phase diagram calculated using the winding number invariant, defined for chiral-
symmetric systems. The values $\nu=\pm1$ in the lower and upper non-trivial area indicate 
that these regions correspond to different topological phases, with one zero energy
mode in each.
}
\end{figure}

The Hamiltonian $H_{\mathrm{BdG}}$, like all Bogoliubov-de Gennes Hamiltonians, is by construction invariant under a particle-hole operation. That is, we can define an antiunitary operator $\mathcal{P}$, such that $\mathcal{P}H_{\mathrm{BdG}}\mathcal{P}^{-1} = -H_{\mathrm{BdG}}$. The action of $\mathcal{P}$ on the original electron operators and on doubled Hilbert space states is
\begin{equation}
\label{eq:phs}
\mathcal{P}c_{is} = c_{is}^\dag,\quad
\mathcal{P}\ve{\chi}_i = (v_{i\uparrow}^*,v_{i\downarrow}^*,u_{i\uparrow}^*,u_{i\downarrow}^*)^T.
\end{equation}
The particle-hole operation maps the positive energy solutions onto their Nambu partners with negative energy.
If the particle-hole symmetric Hamiltonian of a finite system has zero energy modes, they can be cast in the form of eigenstates of $\mathcal{P}$,
\begin{equation}
 \label{eq:majorana-property}
 \mathcal{P}\psi = \psi.
\end{equation}
Inspecting the first relation of (\ref{eq:phs}) shows that (\ref{eq:majorana-property}) is only an equivalent
definition of the Majorana property, usually stated as $\gamma_{\sigma}(\ve{r}) = \gamma^\dag_{\sigma}(\ve{r})$, where $\gamma^\dag$ is the operator creating a particle with spin $\sigma$ at position $\ve{r}$. \\
The presence or absence of Majorana solutions can be predicted from a topological phase diagram, where
different phases correspond to different values of a topological invariant. 
In a system with translational symmetry, such as the bulk of the CNT, the basic quantity determining 
the topological invariant in 1D is $\gamma^-$, the sum of the Berry phases carried by 
all occupied (negative energy) bands, integrated over the Brillouin zone.
Since $\gamma^-$ is gauge-dependent and defined only up to an integer,
another invariant is commonly used, $W=\exp(i2\pi\gamma^-)$, which is gauge-independent.
The particle-hole symmetry in a system with translational invariance is expressed as 
$\mathcal{P}H_{\mathrm{BdG}}(k)\mathcal{P}^{-1} = -H_{\mathrm{BdG}}(-k)$~\cite{chiu:rmp2016,sato:rpp2017},
i.e. the positive energy solutions at momentum $k$ are related to negative energy solutions at momentum $-k$, as
sketched in Fig.~\ref{fig:invariants}(a). This constrains the values which $W$ can take to $\pm 1$, 
i.e. $W$ is of a $\mathbb{Z}_2$ type, associated with the Altland-Zirnbauer D class systems~\cite{chiu:rmp2016,sato:rpp2017}.
$W=+1$ corresponds to the trivial topological phase, while $W=-1$ implies the presence of MQPs at the system boundaries.
The phase diagram calculated for our model nanotube, using the standard Pfaffian technique~\cite{chiu:rmp2016,sau:prb2013}
and the effective model for the bulk bands, is shown in Fig.~\ref{fig:invariants}(b). 
The borders between different phases in the
diagram correspond to $(B_\perp,\mu)$ such that the gap is closed at $k=0$. From our effective four-band model
we find that this occurs at
\begin{eqnarray}
\tilde{\mu}^2 &= &\Delta_{\mathrm{SO}}^2 + 4(\Delta_{KK'}^2 + (\mu_B B_\perp)^2 -\Delta_0^2)
\nonumber \\
& & \pm\sqrt{4\Delta_{KK'}^2((\mu_B B_\perp)^2 - \Delta_0^2)-\Delta_0^2\Delta_{\mathrm{SO}}^2}, \label{eq:gap-closing}
\end{eqnarray}
where $\tilde{\mu}$ is the chemical potential measured from the center of either the \ding{192},\ding{193} or \ding{194},\ding{195}
pair in Fig.~\ref{fig:setup-bulk}(b).
The critical magnetic field is given by $\mu_BB_c=2\Delta_0\Delta_{KK'}/\sqrt{\Delta_{SO}^2 + 4\Delta_{KK'}^2}$.
If we assume that the band pair \ding{192},\ding{193} is independent
of \ding{194},\ding{195}, we can expand (\ref{eq:gap-closing}) around $B_c$, obtaining a simpler formula
$\tilde{\mu}^2 = \Delta_0^2((B_\perp/B_c)^2-1)$. The red lines in Fig.~\ref{fig:invariants}(b)
follow (\ref{eq:gap-closing}), the dashed lines mark the borders of the non-trivial phase obtained 
with the simpler approximated formula.  
The coupling between the band pairs changes visibly the phase diagram -- when the Zeeman energy reaches the
magnitude of the original spin-orbit splitting, it destroys the topological phase. The same phenomenon occurs
in multiband semiconducting nanowires, where the mixing between various transverse modes caused by the Rashba
spin-orbit coupling strongly reduces the non-trivial topological regions in the phase diagram
~\cite{stanescu:jpcm2013,lutchyn:prb2011,lim:prb2012}.\\

As can be seen in Fig.~\ref{fig:BdGspectrum}, the Hamiltonian $H_{\mathrm{BdG}}$ is highly symmetric. In particular,
a unitary operation $C$ can be defined, such that $CH_{\mathrm{BdG}}(k)C^{-1} = -H_{\mathrm{BdG}}(k)$. The operation
$C$ is a so-called chiral symmetry, connecting positive and negative energy solutions at the same momentum $k$, as
sketched in Fig.~\ref{fig:invariants}(c). The MQPs in our system are also eigenstates of $C$.
In systems with this symmetry, the topological invariant $\gamma^-$ has a clear 
interpretation as a winding number, $\gamma^- = \nu/2$~\cite{izumida:prb2017}. The winding number is an integer, i.e. it belongs to $\mathbb{Z}$.
That apparent contradiction with $W\in\mathbb{Z}_2$ is solved when we recall that $W$ was constructed with an extra exponentiation step, which
obliterates the difference between the phases with $\nu=\pm 1$. The phase diagram calculated using the winding number is shown
in Fig.~\ref{fig:invariants}(d), with exactly the same phase boundaries, but
showing clearly that the lower non-trivial region and the upper non-trivial region in fact correspond to different
non-trivial phases. 
Further discussion on this subject can be found in Sec. IV of the Supplemental Material.
\\
With both the particle-hole $\mathcal{P}$ and the chiral $C$ symmetries, the Hamiltonian is also invariant under 
a product of both, i.e. $\tilde{\mathcal{T}}=C\mathcal{P}^{-1}$. This symmetry is antiunitary and commutes with the Hamiltonian, 
$\tilde{\mathcal{T}}H_{\mathrm{BdG}}(k)\tilde{\mathcal{T}}^{-1} = H_{\mathrm{BdG}}(-k)$, similar to the time-reversal. Contrary to the 
true time-reversal operation which in systems with half-integer spin squares to $-1$, here $\tilde{\mathcal{T}}^2=+1$, placing our 
nanotube not in the D, but in the BDI class with the winding number as an integer topological invariant.\\

%

\section{Emergence of  MQPs in finite nanotubes}
\begin{figure}[htp]
\includegraphics[width=\columnwidth]{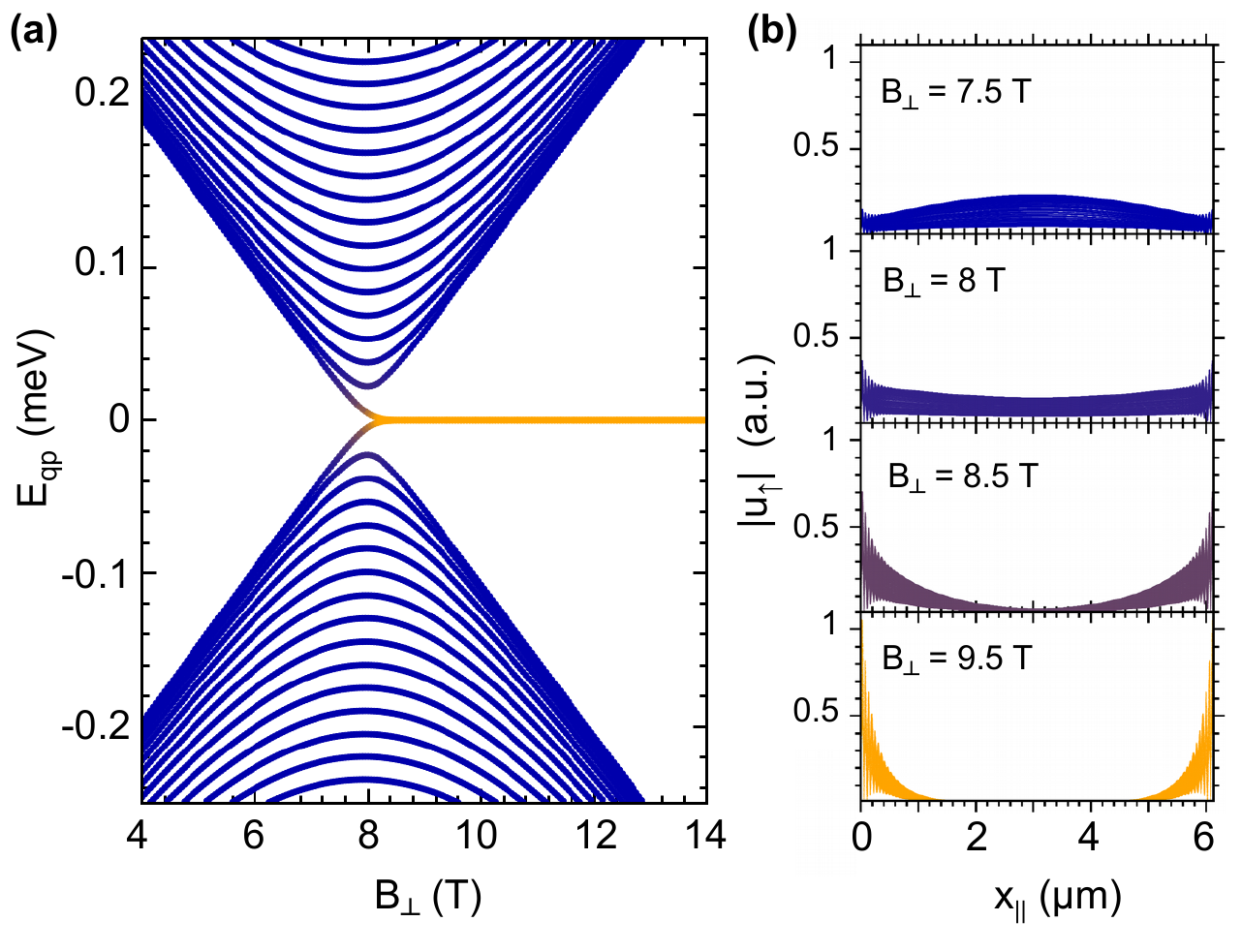}
\caption{\label{fig:phase-transition}
Topological phase transition.
(a) The quasiparticle spectrum of a finite (12,4) nanotube with 4000 unit cells ($L=6.03\,\mu$m), at the chemical potential
$\mu=334.6$~meV for varying magnetic field.
The topological phase transition occurs at $B_c = 8.5$~T, beyond which the lowest energy eigenstate becomes a zero energy mode.
(b) The wave function of the lowest energy mode undergoes a gradual localization with increasing magnetic field. Here only 
the amplitude $|u_\uparrow(x_\parallel)|$ of the spin up particle component, projected onto the direction along the CNT's
axis, is shown. The shape of the remaining components is indistinguishable from that of $|u_\uparrow(x_\parallel)|$
at this scale, which comprises the data points from $N=8.32\cdot10^5$ atoms. The units are arbitrary and the same for all 
wave function plots in this figure.}
\end{figure}

Changing the chemical potential or the strength of the magnetic field can drive the proximitized nanotube across a topological phase
transition, into a regime in which it becomes a topological superconductor. An example of the changes in the 
Bogoliubov-de Gennes spectrum during such a transition is shown in Fig.~\ref{fig:phase-transition}(a), for a 
6~$\mu$m long (12,4) CNT at a fixed chemical potential $\mu=334.6$~meV and varying magnetic field $B_\perp$. 
The energy of the lowest quasiparticle states is further lowered with increasing $B_\perp$, until they become a doubly degenerate
zero energy mode. The degeneracy is artificial, caused by the doubling of degrees of freedom introduced with the Nambu spinor, and the nanotube
de facto hosts only one eigenstate at zero energy. The change in the shape of the quasiparticle wave function associated with the lowest
energy eigenstate is illustrated in Fig.~\ref{fig:phase-transition}(b), showing clearly its increasing localization at the ends of the proximitized
CNT. In the figure only the amplitude $|u_\uparrow(\ve{r})|$ of the particle component with spin up  is shown, the remaining components
$u_\downarrow(\ve{r}),v_\uparrow(\ve{r}), v_\downarrow(\ve{r})$ have profiles which are indistinguishable from $|u_\uparrow(\ve{r})|$ at this scale.
Having a direct access to the particle and hole components of the zero energy mode, we can prove that it indeed has Majorana nature
according to (\ref{eq:majorana-property}).

\begin{figure}[htp]
\includegraphics[width=\columnwidth]{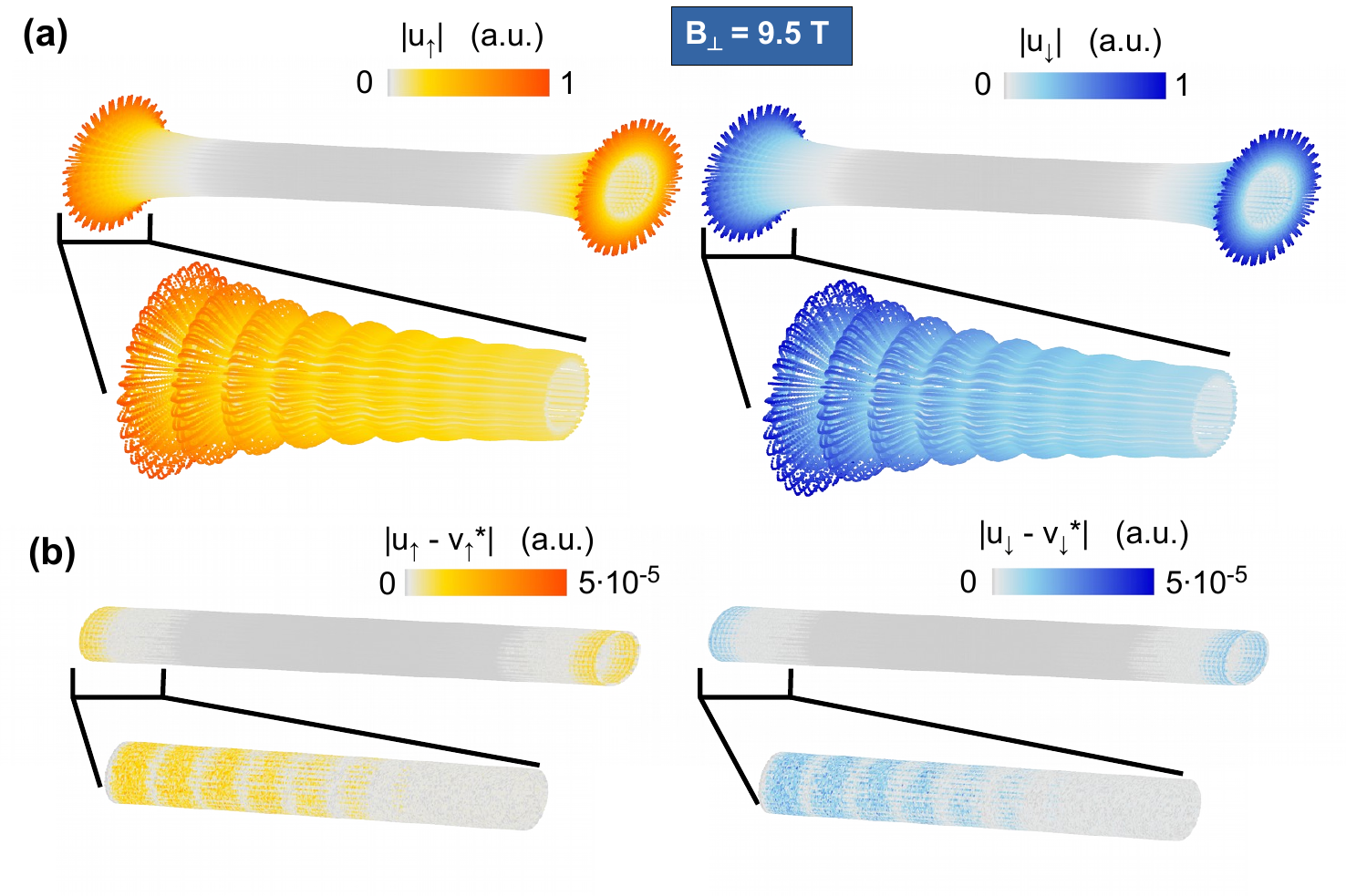}
\caption{\label{fig:majoranas}
Majorana bound states.
(a), The full spatial profile of the the spin up and spin down particle components, $|u_\uparrow(\ve{r})|$ and
$|u_\downarrow(\ve{r})|$. The amplitude of the electronic wave function is shown through both the distance from the nanotube's
surface (light grey) and through the color scale. The wavelength of the oscillations is given by the value of $k_F$ at the
chosen chemical potential.
(b), Spatially resolved amplitude of the {\em difference} between the particle and conjugated hole components for the same spin, 
$|u_\uparrow(\ve{r}) - v^*_\uparrow(\ve{r})|$ and $|u_\downarrow(\ve{r}) - v^*_\downarrow(\ve{r})|$. The distance from the CNT's
surface is scaled in the same way as in {\bf c}, and the color scale is greatly enhanced. Only faint differences are visible,
of the order of $10^{-5}$, which shows the Majorana nature of the zero energy mode.
}
\end{figure}

The spatially resolved wave function of the zero energy mode
at $B_\perp = 9.5$~T is shown in Fig.~\ref{fig:majoranas}(a). The amplitude of spin up and down particle components, $|u_\uparrow(\ve{r})|$ and
$|u_\downarrow(\ve{r})|$, is shown both as the distance from the CNT's surface (grey) at each atomic position and via the
color scale. The wavelength of the oscillations is set by the value of $k_F$ at the chosen chemical potential. The decay length is field-dependent and at $B_\perp = 9.5$ it is $\sim 0.4\,{\mathrm{\mu m}}$. The Majorana nature
of the zero energy mode becomes evident in the Fig.~\ref{fig:majoranas}(b), where the differences between particle and (complex conjugated)
hole component of the wave function for each spin, $|u_\uparrow(\ve{r}) - v^*_\uparrow(\ve{r})|$ and $|u_\downarrow(\ve{r}) - v^*_\downarrow(\ve{r})|$
are shown. They are identical up to the order of $10^{-5}$ of the maximum amplitude, which constitutes a numerical proof that the
zero energy mode fulfills the Majorana condition (\ref{eq:majorana-property}).\\

\section{MQP stability and experimental feasibility}
\begin{figure*}[htp]
\includegraphics[width=\textwidth]{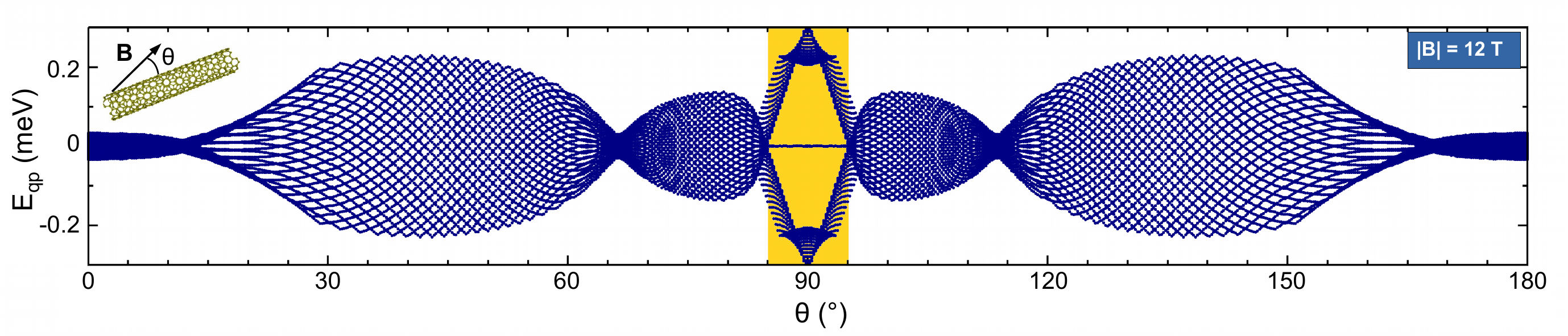}
\caption{\label{fig:stability}
Stability of MQPs with respect to magnetic field alignment.
The thirty two lowest quasiparticle energies as a function of varying angle of the magnetic field, with its 
amplitude fixed at $B = 12$~T. The bulk Hamiltonian is gapped only within the area marked in yellow, 
$85^\circ \leq \theta \leq 95^\circ$. In a finite system a zero energy mode appears throughout this range of $\theta$,
with maximum distance to the other eigenstates at $\theta = 90^\circ$.}
\end{figure*}

The stability of the MQPs against perturbations is crucial for their experimental realization. The techniques for growing carbon nanotubes
are now so advanced that their atomic lattices are nearly perfect~\cite{cao:natmater2005}.
We have analyzed our model nanotube at
several low concentrations of impurities and found that the Majorana mode is remarkably stable (cf. Sec. VI of the Supplemental Material). 
Another factor which has to be taken into account is the precision of alignment of the magnetic field.
The presence of a field component parallel to the nanotube axis gives rise to the Aharonov-Bohm effect. In nanotubes this 
causes a different orbital response in the two valleys, resulting in a removal of the valley
degeneracy~\cite{ajiki:jpsj1993}
and breaking of the chiral symmetry. When the parallel component of the magnetic field reaches a threshold value, the electrons on opposite sides of the $\Gamma$ point no longer have matching momenta and the superconducting correlations become ineffective, yielding a gapless spectrum. The lowest thirty two eigenvalues of the
Bogoliubov-de Gennes spectrum in magnetic field of 12~T amplitude and varying angle $\theta$ with respect to the nanotube axis
are plotted in Fig.~\ref{fig:stability}. 
At this chosen field amplitude the finite system supports a Majorana mode within a range of $\pm5^\circ$ deviation of the field from the perpendicular.
Increasing the field amplitude widens the maximum gap at 90$^\circ$, but the higher value of the parallel component decreases the $\theta$ range in which
the spectrum is gapped. \\
The two major experimental challenges in achieving the formation of MQPs in this setup are the necessity of controlling the chemical potential of the CNT and of applying a large magnetic field without destroying superconducting correlations. Both may be accomplished with the use of 2D transition metal dichalcogenide (TMDC) superconductors, such as NbSe$_2$, with its larger superconducting gap of 1.26~meV~\cite{huang:prb2007}. The superconducting pairing was demonstrated to survive in fields up to 30~T~\cite{xi:natphys2016}, and the thinness of the 2D layer allows the superconductor itself to be gated, together with the CNT in its proximity. 
Thus the goal of tuning a CNT into the non-trivial topological regime, with its attendant Majorana boundary modes, is within the reach of state-of-the-art technology.\\

\acknowledgments{
The authors thank the Deutsche Forschungsgemeinschaft for financial support via 
GRK 1570 and IGK ``Topological insulators'' grants, as well as the JSPS for the KAKENHI Grants (Nos. JP15K05118, JP15KK0147, JP16H01046).
We acknowledge the useful discussions with J. Klinovaja, M. Wimmer and K. Flensberg.
We are grateful to B. Siegert for his advice regarding the numerical calculations.}

\appendix

\section{Tight-binding model} 
The model is constructed for one $p_z$ orbital per atomic site. The hopping matrix elements, taking into account the hybridization between $\sigma$ and $\pi$ orbitals and the spin-orbit coupling induced by the curvature, are given by the formulae in Refs.~\cite{ando:jpsj2000,delValle:prb2011}. In our calculations we chose  $V_{pp}^\sigma = 6.38$~eV and $V_{pp}^\pi = -2.66$~eV after Ref.~\onlinecite{tomanek:prb1988}, and we set the small parameter controlling nanotube's spin-orbit coupling to $\delta_{SO}=3 \cdot 10^{-3}$, similar to $2.8 \cdot 10^{-3}$ measured in Ref.~\onlinecite{jhang:prb2010}. 
The electrostatic potential of the substrate is taken here to be a continuous ridge,  
adding an on-site potential term to the Hamiltonian of the CNT at the atomic sites in the proximity of the superconducting substrate. We have
tested several shapes of this ridge  with similar values of the resulting valley mixing energy scale, $\Delta_{KK'}$.
For all calculations presented here we chose a Gaussian form of $V(\varphi)$, 
\begin{equation}
\label{eq:substrate}
V(\varphi) = V_0\,\exp(-(\varphi-\varphi_0)^2/\Delta\varphi^2),
\end{equation}
where $\varphi$ is the angular coordinate of the nanotube atom, $V_0$ is an arbitrarily chosen maximum height
of the substrate's potential, $\varphi_0$ is the shift between nanotube coordinates and the CNT-S contact line,
and $\Delta\varphi$ controls the sharpness of the potential (more details can be found in Sec. I of the Supplemental Material). In the calculations we assumed $V_0=0.4$~eV, $\varphi_0 = 90^\circ$ and $\Delta\varphi=2.5^\circ$.\\

\section{Effective four-band model}
The Hamiltonian of a CNT in the reciprocal space is obtained using a zone folding technique. The spectrum of the CNT is follows from that of graphene by imposing the periodic boundary conditions on the value of transverse momentum, turning the 2D dispersion of graphene into a series of 1D cuts, which are the CNTs one-dimensional subbands~\cite{saito:1998}. When the curvature of the CNT's lattice is included, it results in both spin-dependent and spin-independent modifications of graphene's dispersion. They are most significant near the Dirac points of the spectrum. In models treating one $p_z$ orbital per atomic site their effects can be incorporated in the dispersion as shifts in both transverse and longitudinal momentum. The low energy electronic spectrum of a CNT in the conduction band for given transverse momentum $k_\perp$ and longitudinal momentum $k$ is then given by
\begin{eqnarray}
 \varepsilon_{\tau\sigma}\left(k_\perp,k\right) & = & \hbar v_{F} \left\{\left(k -\tau K_\parallel + \tau\Delta k_\parallel^{c}\right)^{2}\right.   \label{eq:CNTdispersionFull}
\\
& &+\left. \left(k_{\perp} -\tau K_\perp + \tau\Delta k_{\perp}^{c} + s\Delta k_{\perp}^{\mathrm{SO}}\right)^{2}\right\}^{1/2},\nonumber
 \end{eqnarray}
where $K_\perp,K_\parallel$ are the transverse and longitudinal component of momentum at the Dirac point $K$. The quantum numbers $\tau$ and $s$ stand for the valley ($K:\;\tau=1$,
$K':\;\tau=-1$) and the spin component along the CNT axis $s=\pm 1$.
All quantities in this dispersion are directly related to the hopping integrals across $\pi$ ($V_{pp}^\pi)$ and $\sigma$ bonds ($V_{pp}^\sigma$) in graphene, to nanotube geometry and to carbon's intrinsic spin orbit coupling~\cite{ando:jpsj2000,delValle:prb2011}, and their values and signs may vary, depending on which set of tight-binding parameters is used. The numerical values of those  momentum shifts in our calculations are
$\Delta k_\perp^c = -22.83\,\mu\mathrm{m}^{-1}$, $\Delta k_\parallel^c = 66.62\,\mu\mathrm{m}^{-1}$ and $\Delta k_{\mathrm{SO}}=-2.917\,\mu\mathrm{m}^{-1}$. 
In the case of our (12,4) semiconducting nanotube $K_\parallel=0$ and the lowest energy subbands shown in Fig.~{\bf 1}b have $k_\perp - \tau K_\perp = \tau/3R$. In the following we shorten the notation by setting $k_\perp=K_\perp+1/3R$ and omitting it from the argument of $\varepsilon_{\tau s}(k_\perp,k)$. The spin-orbit splitting $\Delta_{\mathrm{SO}}$ from the main text is then $\Delta_{\mathrm{SO}}=\varepsilon_{K\uparrow}(0)-\varepsilon_{K\downarrow}(0)$. Note that the single-particle energies satisfy the time-reversal conjugation, $\varepsilon_{\tau s}\left(k\right) = \varepsilon_{-\tau-s}\left(-k\right)$.\\
With added valley-mixing induced by the superconducting substrate and in an external perpendicular magnetic field the CNT is described by the following effective Hamiltonian :
\begin{equation}
H = H_{\text{CNT}} + H_{\Delta_{KK^{\prime}}} +  H_{\text{Z}}\text{.}
\label{eq:fullSPHam}
\end{equation}
\noindent
The effective Hamiltonian in second quantization for the CNT including a reference chemical potential $\mu$ is given by 
\begin{equation}
H_{\text{CNT}} - \mu N = \sum_{k,\tau,s}\xi_{\tau s}\left(k\right)c_{k\tau s}^{\dagger}c_{k\tau s}\text{,}
\label{eq:CNTHam}
\end{equation}
\noindent
where $\xi_{\tau s}\left(k\right)$ is the single-particle energy measured with respect to the chemical potential, $\xi_{\tau s}\left(k\right) = \varepsilon_{\tau s}\left(k\right) - \mu$. We model the $k$ dependence of the valley mixing potential  (see Sec. I of the Supplemental Material for details)
by modifying the longitudinal curvature shift and fitting an appropriate constant $\Delta_{KK'}$ to the band structure obtained from the real space calculation. In our case
$\varepsilon_{\tau s}\left(k\right) = \hbar v_{F} \sqrt{\left(k + 0.8 \tau\Delta k_{\parallel}^{c}\right)^{2} + \left(1/3R + \Delta k_{\perp}^{c} + \tau s\Delta k_{\perp}^{SO}\right)^{2}}$. 
The valley-mixing term $H_{\Delta_{KK^{\prime}}}$ couples two electron states at opposite valleys but with the same spin $s$ and becomes
\begin{equation}
H_{\Delta_{KK^{\prime}}} = \sum_{k,s}  \Delta_{KK^{\prime}}c_{k Ks}^{\dagger}c_{k K^{\prime}s} 
+  \Delta_{KK^{\prime}}^{\star}c_{k K^{\prime}s}^{\dagger}c_{k Ks}\text{,}
\label{eq:VMHam}
\end{equation}
with $\Delta_{KK^{\prime}}\in\mathbb{C}$. In our calculations $\Delta_{KK'}$ is real and equal to 2.5~meV. 
The Zeeman energy $H_{\text{Z}}$ due to the perpendicular magnetic field $B_{\perp}$ induces a coupling of electrons with opposite spins and in the same valley
\begin{equation}
H_{Z} = \mu_{B}B_{\perp}\sum_{k,\tau}c_{k\tau\uparrow}^{\dagger}c_{k\tau\downarrow} + c_{k\tau\downarrow}^{\dagger}c_{k\tau\uparrow}\text{,}
\label{eq:ZeeHam}
\end{equation}
i.e. we assume $B_\perp$ to be applied in the $x$ direction, while the $z$ direction runs along the CNT axis.
The eigenstates of the resulting Hamiltonian are then in general linear combinations of all $\tau,s$ eigenstates of the original $H_{\text{CNT}}$. We denote them by  \ding{192},\ding{193},\ding{194},\ding{195}, shown in Fig.~\ref{fig:setup-bulk}b. 

The superconducting correlations induced by proximity are treated in a mean-field approximation. We only consider the case of an on-site pairing potential which is described by the superconducting gap $\Delta_{0}$. Since $\Delta_{0}$ is isotropic in momentum space, our mean-field pairing Hamiltonian has an s-wave gap symmetry. The mean-field Hamiltonian reads
\begin{equation}
\label{eq:pairing}
H_{\text{SC}} = \sum_{k}\Delta_{0}\left(c_{k K\uparrow}^{\dagger}c_{-k K^{\prime}\downarrow}^{\dagger} + c_{k K^{\prime}\uparrow}^{\dagger}c_{-k K\downarrow}^{\dagger} + \mathrm{h.c.}\right)\text{,}
\end{equation}
where we are coupling the corresponding Kramers partners. Introducing the Nambu spinor defined as
\begin{equation*}
\label{eq:NambuEff}
\Psi^\dag = (c_{kK\uparrow}^\dag, c_{kK\downarrow}^\dag, c_{kK'\uparrow}^\dag, c_{kK'\downarrow}^\dag, c_{-kK'\downarrow},c_{-kK'\uparrow},c_{-kK\downarrow}, c_{-kK\uparrow}),
\end{equation*}
we obtain the Bogoliubov-de Gennes (BdG) Hamiltonian
\begin{equation}
\label{eq:four-band-BdG}
\mathcal{H}_{\mathrm{BdG}}(k) =  \begin{pmatrix}
                               H(k) & \Delta \\
                               -\Delta & -H(-k)
                              \end{pmatrix},
\end{equation}
with
\begin{equation}
\label{eq:sppHam}
H(k) = \begin{pmatrix}
        \xi_{K\uparrow}(k) & \mu_B B_\perp & \Delta_{KK'} & 0 \\
        \mu_B B_\perp & \xi_{K\downarrow}(k) & 0 & \Delta_{KK'} \\
        \Delta_{KK'} & 0 & \xi_{K'\uparrow}(k) & \mu_B B_\perp \\
        0 & \Delta_{KK'} & \mu_B B & \xi_{K'\downarrow}(k)
       \end{pmatrix}, 
\end{equation}
and
\begin{equation*}
 \Delta = \begin{pmatrix}
           -\Delta_0 & & & \\
           & \Delta_0 & & \\
           & & -\Delta_0 & \\
           & & & \Delta_0 \\
          \end{pmatrix}.
\end{equation*}
The single particle energies are defined with respect to the chemical potential $\mu$, as in \eqref{eq:CNTHam}. 
The resulting p-wave and s-wave components of the pairing Hamiltonian in the eigenbasis of the \ding{192},\ding{193},\ding{194} and \ding{195}
are shown in Fig.~\ref{fig:pairings} and discussed further in Sec. II of the Supplemental Material. A detailed analysis of
the superconducting pairing, also within an effective two-band model can be found in Sec. III of the Supplemental Material.\\

\section{Gap closing condition}
The Hamiltonian $\mathcal{H}_{\mathrm{BdG}}(k)$ has a chiral symmetry,
i.e. there exists a unitary operator $C$ such that $C\mathcal{H}_{\text{BdG}}\left(k\right)C^{-1} = -\mathcal{H}_{\text{BdG}}\left(k\right)$. In the basis in which the operator $C$ is diagonal (details can be found in the next section), the BdG Hamiltonian is given by
\begin{equation*}
\mathcal{H}^{C}_{\text{BdG}}\left(k\right) = \begin{pmatrix}
0 & D\left(k\right) \\
D^{\dagger}\left(k\right) & 0
\end{pmatrix}\text{,}
\label{eq:BdGChiralBasis}
\end{equation*}
where $D(k) = H(k) - i\Delta$. In order to obtain the gap closing condition we square the BdG Hamiltonian in chiral basis, which yields
\begin{equation*}
\left(\mathcal{H}^{C}_{\text{BdG}}\left(k\right)\right)^{2} = \begin{pmatrix}
D\left(k\right)D^{\dagger}\left(k\right) & 0 \\
0 & D^{\dagger}\left(k\right)D\left(k\right)
\end{pmatrix}\text{.}
\end{equation*}
This matrix has zero energy eigenvalues at the $\Gamma$ point if $\det\left(D\left(k = 0\right)D^{\dagger}\left(k = 0\right)\right) = \det\left(D^{\dagger}\left(k = 0\right)D\left(k = 0\right)\right) = 0$. From this we obtain the exact gap closing condition at the $\Gamma$ point,
given by Eq.~\eqref{eq:gap-closing}.\\
\section{Topological invariants. }
The symmetries of the BdG Hamiltonian ~\eqref{eq:four-band-BdG} can be expressed in terms of Pauli matrices, denoted by $\pi$ in the particle-hole (Nambu) subspace, by $\tau$ in the valley subspace and by $s$ in the spin subspace. The particle-hole symmetry operator $\mathcal{P}$, such that  $\mathcal{P}\mathcal{H}_{\mathrm{BdG}}(k)\mathcal{P}^{-1} = -\mathcal{H}_{\mathrm{BdG}}(-k)$, is given by $\mathcal{P} = \pi_x\otimes\tau_x\otimes s_x\,\mathcal{K}$, 
where $\tau_0$ and $s_0$ are the identities in their respective subspaces and $\mathcal{K}$ denotes the operator of the complex conjugation. The  Hamiltonian $H_{\mathrm{BdG}}$ has also a chiral symmetry, i.e. it fulfills $C \mathcal{H}_{\mathrm{BdG}}(k)C^{-1} = -\mathcal{H}_{\mathrm{BdG}}(k)$ with a unitary operator $C$. The operator is given by $C = \pi_y\otimes\tau_0\otimes s_0$.
The presence of those two symmetries implies that there exists a third one, which we call $\tilde{\mathcal{T}}=C\mathcal{P}^{-1}$ and which fulfills $\tilde{\mathcal{T}}\mathcal{H}_{\mathrm{BdG}}(k)\tilde{\mathcal{T}}^{-1} = \mathcal{H}_{\mathrm{BdG}}(-k)$. Its expression in this basis is
$ \tilde{\mathcal{T}} = -i\pi_z\otimes\tau_x\otimes s_x\, \mathcal{K}$.
The operation $\tilde{\mathcal{T}}$ squares to $+1$, hence it is clear that it is not the time reversal symmetry of a spin-1/2 system. The fact that it is diagonal in the Nambu space implies that already the non-superconducting Hamiltonian $H(k)$~\eqref{eq:sppHam} is invariant under a restricted $\tilde{\mathcal{T}}_{\mathrm{red}}=\tau_x\otimes s_x\mathcal{K}$, which is indeed the case and reflects a physical symmetry of the system. It is the symmetry of rotation with respect to an axis {\em perpendicular} to the CNT, which exchanges both the valley, longitudinal momentum and spin. 
It also exchanges the sublattices, which accounts for its $\mathcal{K}$ component. 
If, and only if, the magnetic field is also applied perpendicular to the CNT axis, the non-superconducting Hamiltonian is invariant under $\tilde{\mathcal{T}}_{\mathrm{red}}$.\\
{\em Pfaffian $(\mathbb{Z}_2)$ invariant. }
In systems with particle-hole symmetry the topological invariant $W$ can be evaluated using the representation of the Hamiltonian in the
Majorana basis, i.e. the basis of eigenstates of $\mathcal{P}$~\cite{chiu:rmp2016}, obtained by a transformation $U_M$,
$\mathcal{H}_M(k) = U_{\mathrm{M}} \mathcal{H}_{\mathrm{BdG}}(k) U_{\mathrm{M}}^\dagger$.
We can define a matrix $X$ by $i X(k) = \mathcal{H}_M(k)$.
At the time reversal invariant momenta $k=0, \pi/a$, $X(k)$ is a real and skew symmetrix matrix, $X(k) = -[X(k)]^T$.
The topological invariant $W$ can then be expressed through the Pfaffian of $X$ at $k=0,\pi/a$~\cite{chiu:rmp2016},
$W = \mathrm{sgn} \left\{ \mathrm{Pf} [X(\pi)]\;\mathrm{Pf}[X(0)] \right\}
  = \pm 1$, which is of a $\mathbb{Z}_2$ type.
For our system, the unitary matrix $U_M$ is given by
\begin{equation*}
  U_{\mathrm{M}}
=
  \frac{1}{\sqrt{2}}
  \left(
  \begin{array}{cc}
    1 & 1 \\
    -i & i 
  \end{array}
  \right) \otimes \tau_x \otimes s_x.
\end{equation*}
At time reversal invariant momenta $k=0, \pi$, $X(k)$ has the particularly simple form,
\begin{equation*}
  X(k) 
  = 
  \left(
  \begin{array}{cc}
    0 & H(k) + \Delta_M \\
    - \left( H(k) + \Delta_M \right) & 0
  \end{array}
  \right).
\end{equation*}
Then the Pfaffian is calculated as $\mathrm{Pf}[X(k)] = \mathrm{det} \left[ H(k) + \Delta_M\right]$.
We calculated the topological phase diagrams with the $W$ invariant numerically, 
assuming $\mathrm{sgn} \mathrm{Pf}[X(\pi)] = +1$ thus checking only for the band inversion at $k=0$.  \\
{\em Winding number $(\mathbb{Z})$ invariant.}
Since the BdG Hamiltonian has the chiral symmetry $\{C, \mathcal{H}_{\mathrm{BdG}}\} = 0$, one can introduce the winding number
$\nu
  = -\frac{1}{4 \pi i} 
  \int_{\mathrm{BZ}}
  \mathrm{Tr} \left[ C \mathcal{H}_{\mathrm{BdG}}^{-1}(k) \partial_k \mathcal{H}_{\mathrm{BdG}}(k) \right]
  \label{eq:w_def}
$
as a 1D topological invariant~\cite{wen:npb1989,sato:prb2011}.
The identity with another definition of the winding number, which uses
a flat band Hamiltonian~\cite{schnyder:prb2008}, is proven in
Appendix C1 in Ref. \onlinecite{izumida:prb2017}.
Let us consider the unitary transformation
\begin{equation*}
  U_c = 
  \frac{1}{2} 
  \left(
  \begin{array}{cc}
     1 + i & 1 + i \\
    -1 + i & 1 - i
  \end{array}
  \right)\otimes\tau_0\otimes s_0,
  \label{eq:U_pi}
\end{equation*}
which rotates the Pauli matrices for the particle-hole basis as
$U_c^\dagger {\pi}_x U_c = {\pi}_y$,
$U_c^\dagger {\pi}_y U_c = {\pi}_z$,
$U_c^\dagger {\pi}_z U_c = \pi_x$.
Correspondingly, the Hamiltonian in Eq.~\eqref{eq:four-band-BdG} takes
an off-diagonal form,
\begin{equation*}
  \mathcal{H}_c(k) 
  = 
  U_c^\dagger \mathcal{H}_{\mathrm{BdG}}(k) U_c
  = 
  \left(
    \begin{array}{cc}
      0 & H(k)-i\Delta \\
      (H(k) -i\Delta)^\dag & 0
    \end{array}
    \right).
  \label{eq:H_BdG_bulk}
\end{equation*}
Because the chiral operator is transformed as $C_c = U_c^\dagger C U_c = \pi_z$, the winding number is
written as
$ \nu = \frac{1}{2 \pi} \int_{\mathrm{BZ}} dk \partial_k \arg \det (H(k)-i\Delta).
  \label{eq:w_int_arg_det_h}$
The topological invariant $\nu_l$ for the band $l$ can be shown to be $\mathbb{Z} \ni \nu_l = 2\gamma_l$, 
if $\gamma_l$ is calculated in the basis of chiral symmetry eigenstates. Therefore $W = \exp(i\pi \sum_{l} \nu_{l}) = \pm 1$.

%


\end{document}